\documentclass[aps,pra,longbibliography,preprint,floatfix]{revtex4-1}

\usepackage{graphicx}
\graphicspath{{figs/}}
\usepackage{natbib}

\usepackage{amssymb, amsmath, amsfonts, wasysym, bm} 
\usepackage{color}          
\usepackage{float}          
\usepackage{mathrsfs}       
\usepackage{natbib}         
\usepackage{siunitx}        
\usepackage{subfigure}         
\usepackage{todonotes}      
\usepackage{units}          
\usepackage{url}
\usepackage{bigints}

\usepackage[colorlinks=true, 
            linkcolor=blue,
            urlcolor=blue,
            citecolor=blue,
            final,
            hypertexnames=false]{hyperref}
\usepackage{cleveref}

\def\etal{{\it et al.~}}
\newcommand{\vvvert}{|\kern-1pt|\kern-1pt}

\begin{document}
\title[Lyapunov Exponets in Isotropic Turbulence]{Scaling of Lyapunov
  Exponents in Homogeneous Isotropic Turbulence}

\author{Prakash Mohan}
  \email{prakash@ices.utexas.edu}
\affiliation{Institute for Computational Engineering and Sciences,\\ 
The University of Texas at Austin, Austin, Texas 78712, USA}

\author{Nicholas Fitzsimmons
  \footnote{Current affiliation: Applied Research Laboratories,
  the University of Texas at Austin}}
\email{nfitz@arlut.utexas.edu}
\affiliation{Institute for Computational Engineering and Sciences,\\ 
The University of Texas at Austin, Austin, Texas 78712, USA}

\author{Robert D. Moser}
\email{rmoser@ices.utexas.edu}
\affiliation{Institute for Computational Engineering and Sciences,\\ 
The University of Texas at Austin, Austin, Texas 78712, USA}
\affiliation{Department of Mechanical Engineering,\\
The University of Texas at Austin, Austin, Texas 78712, USA}

\begin{abstract}
Lyapunov exponents measure the average exponential growth rate of
typical linear perturbations in a chaotic system, and the inverse of
the largest exponent is a measure of the time horizon over which the
evolution of the system can be predicted. Here, Lyapunov exponents are
determined in forced homogeneous isotropic turbulence for a range of
Reynolds numbers. Results show that the maximum exponent increases
with Reynolds number faster than the inverse Kolmogorov time scale,
suggesting that the instability processes may be acting on length and
time scales smaller than Kolmogorov scales. Analysis of the linear
disturbance used to compute the Lyapunov exponent, and its
instantaneous growth, show that the instabilities do, as expected, act
on the smallest eddies, and that at any time, there are many sites of
local instabilities.
\end{abstract}

\maketitle

\section{Introduction}
One of the defining characteristics of turbulence is that it is
unstable, with small perturbations to the velocity growing
rapidly. Indeed, turbulent flows in closed
domains appear to be chaotic dynamical systems \cite{keefe1992dimension}. The result is
that the evolution of the detailed turbulent fluctuations can only be
predicted for a finite time into the future, due to the exponential
growth of errors. In a chaotic system, this prediction horizon is
inversely proportional to the largest Lyapunov exponent of the
system, which is the average exponential growth rate of typical linear
perturbations. The maximum Lyapunov exponent $\bar\gamma$ is
commonly used to characterize the chaotic nature of a dynamical
system \cite{eckmann1985ergodic}. In a turbulent flow, the maximum Lyapunov
exponent is thus a measure of the strength of the instabilities that
underlie the turbulence, and its inverse defines the time scale over
which the turbulence fluctuations can be meaningfully predicted.

Lyapunov exponents in chaotic fluid flows have been estimated
experimentally since the work of Swinney \cite{wolf1985determining}, using indirect
methods. In numerical simulations, however, Lyapunov exponents can be
determined directly by computing the evolution of linear
perturbations. This has been done for weakly turbulent Taylor Couette
flow \cite{vastano1991short} and very low Reynolds number planar Poiseuille flow
\cite{keefe1992dimension}. Remarkably, to the authors' knowledge, Lyapunov
exponents have not been determined for isotropic turbulence, a
shortcoming corrected in this paper.

Homogeneous isotropic turbulence is an idealized turbulent flow that
has been extensively studied both experimentally \cite{mydlarski1996onset,comte1971simple,comte1966use} and using
numerical simulations \cite{rogallo1981numerical, vincent1991satial, jimenez1993structure, chen1992high}. It is valuable as a model for the
small scales of high Reynolds number turbulence away from walls
\cite{durbin2011statistical}. It has been speculated that in isotropic turbulence, the
maximum Lyapunov exponent scales with the inverse Kolmogorov time
scale \cite{crisanti1993intermittency}, suggesting that the dominant instabilities occur at
Kolmogorov length scales as well.
If true, then a study of the maximum Lyapunov exponent and
the associated instabilities in homogeneous isotropic turbulence will
be applicable to a wide range of flows.

This paper focuses on how the maximum Lyapunov exponent and hence the
predictability time horizon scale with Reynolds number and
computational domain size of a numerically simulated homogeneous
isotropic turbulence. The speculation that $\bar\gamma$ should scale
as the inverse Kolmogorov time scale $\tau_\eta$ \cite{crisanti1993intermittency} is in
agreement with an estimate from a shell model~\cite{aurell1996growth}. However,
this scaling has not been directly tested in direct numerical
simulations.

In addition, in the process of computing the maximum Lyapunov exponent
in a direct numerical simulation, one necessarily computes the linear
disturbance that is most unstable (on average). This can be used in
the short-time Lyapunov exponent analysis, as introduced in
\cite{vastano1991short}, to characterize the nature of the instabilities. This
will be pursued here for isotropic turbulence.

The remainder of this paper includes a brief review of Lyapunov
exponents and how they are computed in numerical simulations
(section~\ref{sec:lyapunov}) followed by a description of the direct
numerical simulations studied here (section~\ref{sec:dns}). The
results of a scaling study of the Lyapunov exponents are given in
section~\ref{sec:scaling}, and a short-time Lyapunov exponent analysis
is presented in section~\ref{sec:short}, followed by concluding
remarks in section~\ref{sec:conclude}.

\section{Lyapunov Exponent Analysis}
\label{sec:lyapunov}
Two important characteristics of chaotic dynamical systems for
the purposes of the current study are that 1) solutions evolve toward a
stable attractor, and 2) solution trajectories on the attractor are
unstable so that near-by trajectories diverge exponentially. The rate
of this exponential divergence is characterized by the Lyapunov
exponents, whose characteristics are recalled briefly here. Further
details can be found in \cite{vastano1991short}. In addition, the use of Lyapunov
exponents in the analysis presented in the paper is described.

\subsection{Evolution of Linear Perturbations}
Consider a solution trajectory $u(t)$ of a chaotic system.  The
solution will evolve toward an attracting set in phase space (the
attractor); in turbulence this corresponds to the solution evolving to
a statistically stationary state. Let $u(t_0)$ at some arbitrary
starting time $t_0$ be on the attractor, and consider an infinitesimal
perturbation $\delta u(t_0)$ of the solution at time $t_0$, and its
evolution in time. The Lyapunov exponents describe the growth or decay
of the magnitude of $\delta u$. In particular, the
multiplicative ergodic theorem \cite{oseledec1968multiple} implies that the limit
\begin{equation}
\bar\gamma=\lim_{t\rightarrow\infty}\frac{1}{t}\log\left(\frac{\|\delta
  u(t)\|}{\|\delta u(t_0)\|}\right)
\end{equation}
exists and $\bar\gamma$ is called a Lyapunov exponent. There is a
spectrum of possible Lyapunov exponents, depending on the solution
$u(t_0)$ and the perturbation $\delta u(t_0)$ at the starting time.
However, for almost all $\delta u(t_0)$, $\bar\gamma=\gamma_1$ the
largest Lyapunov exponent, and, due to round-off error and
other sources of noise, in practical computations, $\bar\gamma=\gamma_1$
for all $\delta u(t_0)$. Furthermore, the Lyapunov spectrum
($\gamma_1>\gamma_2>\gamma_2>\cdots$) does not depend on
$u(t_0)$; it is instead a property of the dynamical
system. See the review by Eckmann \etal\cite{eckmann1985ergodic} for an
introduction to the theory.

In addition, in practical computations as discussed above, we expect that in
the limit $t\rightarrow \infty$
\begin{equation}
\frac{\delta u(t)}{\|\delta u(t)\|}\rightarrow \overline{\delta
  u}(u(t))\qquad\mbox{and}\qquad \frac{1}{\|\delta
    u(t)\|}\frac{d\|\delta u(t)\|}{dt}\rightarrow \gamma'(u(t))
\end{equation}
where $\overline{\delta u}$ and $\gamma'$ depend only on the solution
at $t$, and not on the starting conditions $u(t_0)$ and $\delta
u(t_0)$. The perturbation $\overline{\delta u}$ is the disturbance
that grows most rapidly in the long run, growing at the average exponential
rate $\bar\gamma$. It is defined by the fact that it's long-time average
growth rate forward in time is $\bar\gamma$ and when the evolution is
backward in time the long-time average growth rate is $-\bar\gamma$. The
short-time Lyapunov exponent $\gamma'$ is simply the instantaneous
exponential growth rate of $\overline{\delta u}$.

Because $\gamma'$ and $\overline{\delta u}$ depend only on the
solution at the current time, they can be used as diagnostics for the
instabilities responsible for a system being chaotic. In
particular, when $\gamma'$ is large, the underlying system is
particularly unstable, and at that time the Lyapunov disturbance
$\overline{\delta u}$ is rapidly growing. Thus by seeking out times
when $\gamma'$ is large, and by analyzing the solution $u$ and the
Lyapunov disturbance $\overline{\delta u}$ at that time, we can
characterize the important instabilities. This is the short-time
Lyapunov exponent analysis described by Vastano \&
Moser \cite{vastano1991short}.

In this paper we will be concerned with the scaling of $\bar\gamma$ with
Reynolds number and with the chaotic instabilities revealed by
short-time Lyapunov exponent analysis.

\section{Simulations}
\label{sec:dns}
To simulate the base field, we solve the three-dimensional
incompressible Navier-Stokes equations on a cube of dimension
$L=2\pi$, with periodic boundary conditions, to obtain a
computational approximation of homogeneous isotropic turbulence.
Turbulence is maintained by introducing a forcing term to the
Navier-Stokes equations which only acts at large scales. The forcing
formulation is described in section~\ref{sec:negViscos}.  The
Navier-Stokes equations are solved using a Fourier-Galerkin spatial
discretization with $N$ modes in each direction, and the vorticity
formulation of Kim \etal\cite{kim1987turbulence}.
This formulation has the
advantage of exactly satisfying the continuity constraint while eliminating
the pressure term. A low-storage explicit third-order Runge-Kutta scheme \cite{spalart1991spectral} is used for time evolution.
The simulations are performed using a modified version of the channel flow code
PoongBack~\cite{lee2013petascale,lee2014experiences}.

To compute the Lyapunov exponents, we compute
the growth rate of a linear perturbation added to the base field. This
perturbation satisfies the linearized Navier-Stokes equations:
\begin{equation} \label{lyapNS}
      \frac{\partial \delta u_i}{\partial t} + \frac{ \partial}{\partial
       x_j}(u_j \delta u_i + \delta u_j u_i) = -\frac{\partial
        \delta p}{\partial x_i} + \frac{1}{Re}\nabla^2 \delta u_i
\end{equation}
\begin{equation}
      \partial_i \delta u_i = 0,
\end{equation}
where $u_i$ is the base field and $\delta u_i$ is the disturbance
field. The disturbance equations are solved using the same numerical
scheme as the Navier-Stokes equations.  Note that the forcing is applied only to the base field and
not the perturbation.
The implementation of both the base and disturbance field solvers were
verified using the method of manufactured solutions.

\subsection{Forcing} \label{sec:negViscos}
The goal of the forcing is to inject energy into the
large-scale turbulence so that the isotropic turbulence will be
stationary.  Forcing is applied to Fourier modes with wavenumber
magnitudes in a specified range, and is designed to produce a
specified rate of energy injection (forcing power), which, when the
system is stationary, will be the dissipation rate. By specifying the
wavenumber range being forced, forcing power and viscosity,
the integral scale, turbulent kinetic energy and Reynolds number can
be controlled.

The energy injection is accomplished by the introduction of a forcing
term $f_i$ to the Navier Stokes equations:
\begin{align} \label{nvforceNS}
      \frac{\partial u_i}{\partial t} + \frac{\partial u_iu_j}{\partial x_j}= -\frac{\partial p}{\partial x_i} + \frac{1}{Re}\nabla^2 u_i + f_i \\
       \partial_iu_i = 0.
\end{align}
Following \cite{spalart1991spectral}, in the Fourier spectral method used here, the Fourier transform of the
forcing $\hat f_i$ is specified in terms of the velocity Fourier
transform $\hat u_i$ as
\begin{equation}
  \hat{f}_i(\mathbf{k}) = \alpha |\mathbf{k}|^2 \hat{u}_i(\mathbf{k}).
\end{equation}
Given that $u_i$ is a Navier-Stokes solution, $f_i$ is guaranteed to be divergence-free. The coefficient $\alpha$ in
the above is determined as a function of time so that the forcing power is the target
dissipation rate $\epsilon_T$. Since the forcing is applied only
to a range of wavenumbers, this yields
\begin{equation}
  \alpha =
  \epsilon_T
  \begin{cases}
    {\displaystyle
    \epsilon_T\left(\sum_{k_{f_{\text{min}}}\leq|\mathbf{k}| \leq
      k_{f_{\text{max}}}}
    |\mathbf{k}|^2\hat{u}^*_i(\mathbf{k})\hat{u}_i(\mathbf{k})\right)^{-1}}
    & k_{f_{\text{min}}}\leq|\mathbf{k}|\leq k_{f_{\text{max}}}\\
    0 & \text{otherwise}
  \end{cases}
\end{equation}
where $\cdot^*$ denotes the complex conjugate, and
$k_{f_{\text{min}}}$ and $k_{f_{\text{max}}}$ are the bounds on the
range of wavenumbers being forced. In the Fourier transform of the
Navier-Stokes equations, the viscous term has the same structure as
$f_i$, so this forcing can be interpreted as a negative viscosity
acting in the specified wavenumber range. The combined forcing and
viscous term is then $-(\nu-\alpha)|\mathbf{k}|^2\hat
u_i(\mathbf{k})$. In the numerical solution of the Navier-Stokes
equations, this combined term is treated in the same way as the
viscous term would be. Note that $f_i$ is just a nonlinear function of
$u_i$, so there is no externally imposed stochasticity.

\begin{table}[t]
    \begin{center}
        \begin{tabular}{| r | r | r | r | r | r | r | r |}
        \hline
        Case & $k_{f_{\text{min}}}$ &
        $k_{f_{\text{max}}}$ & \multicolumn{1}{|c|}{$\nu$} & $N$ & $\mathcal{L}$ &
        $Re_\lambda$ & $T_{\rm avg}q/\cal{L}$\\ \hline
        1 & 0 & 2 & 0.0235 & 64  & 1.43  &  37.92 & 455.2 \\ \hline
        2 & 0 & 2 & 0.0113 & 96  & 1.58  &  58.34 & 123.8 \\ \hline
        3 & 0 & 2 & 0.0056 & 128 & 1.67  &  85.68 & 118.0 \\ \hline
        4 & 0 & 2 & 0.0038 & 192 & 1.70  & 106.33 &  51.2 \\ \hline
        5 & 0 & 2 & 0.0026 & 256 & 1.77  & 130.43 &  51.3 \\ \hline
        6 & 0 & 2 & 0.0010 & 512 & 1.82  & 211.76 &  69.5 \\ \hline
        7 & 2 & 4 & 0.0093 & 128 & 0.71  &  37.74 & 277.1 \\ \hline
        8 & 4 & 8 & 0.0037 & 256 & 0.35  &  37.31 &  72.1 \\ \hline
        \end{tabular}
        \caption{Parameters defining the eight direct numerical simulations
          performed to study Lyapunov exponent scaling. Values of
          $\mathcal{L}$ are quoted in  units in which the domain size
          is $2\pi$, and averaging times are normalized by eddy
          turnover time.}
        \label{tab:cases}
    \end{center}
\end{table}

\subsection{Simulation Cases}
To investigate the scaling of the maximum Lyapunov exponent $\bar\gamma$ with
both Reynolds number and the ratio of the computational domain size $L$ to the
integral scale $\mathcal{L}$, eight simulations were performed. These are
summarized in table~\ref{tab:cases}. To study the scaling of $\bar\gamma$ with
Reynolds number, six cases where simulated with the same forcing wavenumber
range and $\epsilon_T$. This resulted in approximately the same integral scale
in each case. The Reynolds number was manipulated by changing the viscosity. To
study the potential variation of $\bar\gamma$ with domain size normalized by
integral scale, the domain size was kept fixed at $2\pi$ and the integral scale
was changed by adjusting the forced wavenumber range, while keeping the
Reynolds number approximately fixed. In all cases  $k_{\text{max}}\eta>1$,
where $k_{\text{max}}$ is the maximum resolved wavenumber, and $\eta$ is the
Kolmogorov scale. In a refinement study, this was found to be sufficient to
obtain resolution independent values of $\bar\gamma$.

For each case, the simulations were run until the base solution became
statistically steady and then the statistics were gathered by time
averaging over a period $T_{\rm avg}$ as reported in
table~\ref{tab:cases}. The simulation was confirmed to be
stationary by verifying the convergence of the viscous dissipation to
$\epsilon_T$ and the statistical convergence rates of $q^2$ and $\bar\gamma$.

\section{Scaling of Lyapunov Exponents}
\label{sec:scaling}
Of primary concern here is the dependence of the maximum Lyapunov
exponent on the Reynolds number and on the domain size. To address
this, the maximum Lyapunov exponent $\bar\gamma$, the integral scale
($\mathcal{L}$) and Reynolds number based on the Taylor micro-scale
($Re_\lambda$) are needed, along with their uncertainties. Based on
the assumption of isotropy, the latter two were determined to be
$\mathcal{L}=0.15q^3/\epsilon$ and
$Re_\lambda=q^2\sqrt{5/(3\epsilon\nu)}$ \cite{pope2001turbulent}. Thus the two
statistical quantities that need to be computed from the DNS are
$\bar\gamma$ and $q^2$. Both are determined as a time average over
averaging time $T_{\rm avg}$ (see table~\ref{tab:cases}), and the
standard deviations $\sigma$ of the uncertainty due to finite averaging
time were determined using the technique described by
\cite{Oliver2014}. The values of $\bar\gamma$, $q^2$ and their
standard deviations are given in table~\ref{tab:data}.  The standard
deviations of the derived quantities $\mathcal{L}$ and $Re_\lambda$
are determined simply as
$\sigma_{\mathcal{L}}=(0.225q/\epsilon)\sigma_{q^2}$ and
$\sigma_{Re_\lambda}=\sqrt{5/(3\epsilon\nu)}\sigma_{q^2}$, where for
$\sigma_{\mathcal{L}}$ it is assumed that $\sigma_{q^2}/q^2\ll 1$. Note
that since for each simulation $\epsilon$ and $\nu$ are specified,
there is no uncertainty in their values.

\begin{table}[t]
  \begin{center}
    \begin{tabular}{| r | r | r | r | r |}
    \hline
 case & $q^2$ & $\sigma_{q^2}$ & $\bar\gamma\tau_\eta$ &
 $\sigma_{\bar\gamma}\tau_\eta$\\\hline
 1 & 4.51 &  0.107 & 0.0922 & 0.0038 \\ \hline
 2 & 4.80 &  0.160 & 0.1075 & 0.0046 \\ \hline
 3 & 4.99 &  0.075 & 0.1177 & 0.0032 \\ \hline
 4 & 5.05 &  0.044 & 0.1231 & 0.0040 \\ \hline
 5 & 5.19 &  0.046 & 0.1304 & 0.0034 \\ \hline
 6 & 5.28 &  0.084 & 0.1599 & 0.0048 \\ \hline
 7 & 2.82 &  0.008 & 0.0941 & 0.0019 \\ \hline
 8 & 1.76 &  0.001 & 0.0945 & 0.0021 \\ \hline
 \end{tabular}
\caption{Values of $q^2$ and the maximum Lyapunov exponent
  $\bar\gamma$, along with the standard deviation ($\sigma$) of the
  sampling uncertainty. Values of $q^2$ are quoted in units in which
  the domain size is $2\pi$ and $\epsilon=1$, and $\bar\gamma$ is
  normalized by the Kolmogorov time scale $\tau_\eta$.}
\label{tab:data}
  \end{center}
\end{table}
\begin{figure}
\begin{center}
    \subfigure[] {\includegraphics[width=0.5\linewidth]{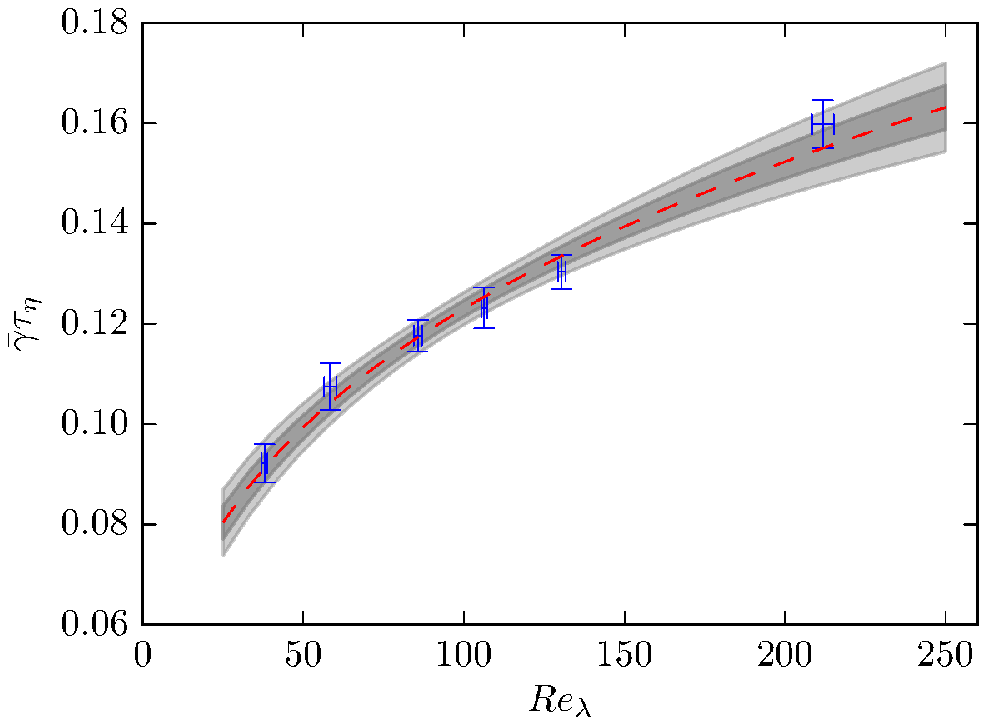}}%
    \subfigure[] {\includegraphics[width=0.5\linewidth]{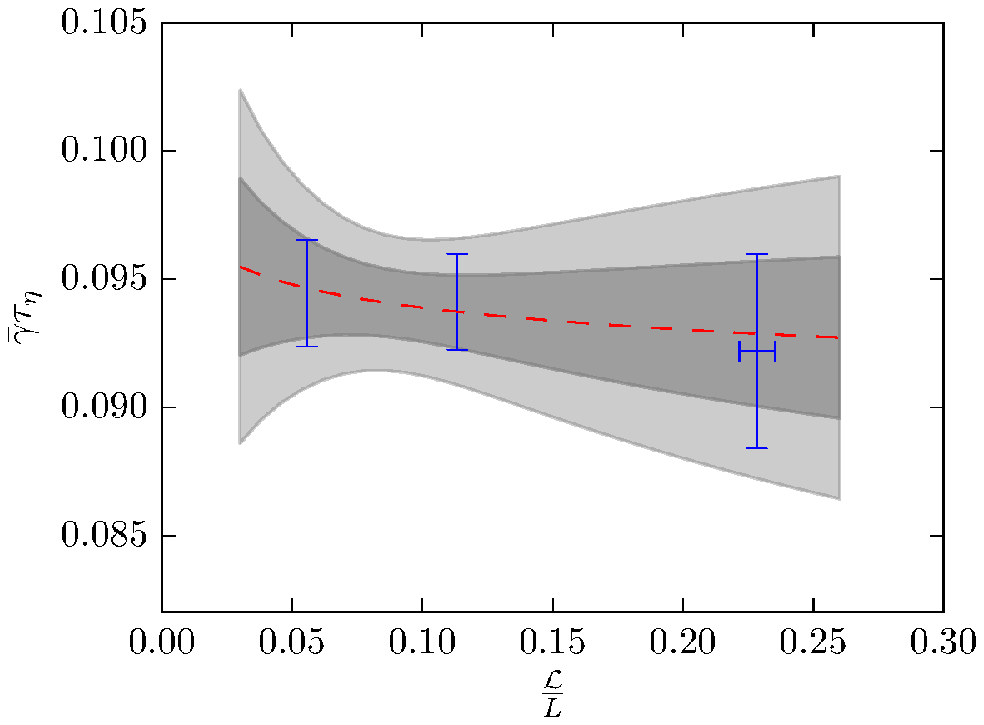}}
\end{center}
\caption{Dependence of the Lyapunov exponent scaled in Kolmogorov
  units ($\bar\gamma\tau_\eta$) on (a) the Taylor scale Reynolds
  number $Re_\lambda$ and (b) the ratio of the integral scale to the
  domain size $\mathcal{L}/L$, from the data in
  table~\ref{tab:data}. The error bars on the data (in blue) represent one
  standard deviation. Also shown are the outputs of the models~(\ref{eq:models})
  (in red) calibrated with Bayesian inference, with the dark and light gray
  shading representing variations of one and two standard deviations respectively.}
\label{fig:dependence}
\end{figure}

The dependence of the maximum Lyapunov exponent in Kolmogorov units on
Reynolds number is shown in figure~\ref{fig:dependence}, including
uncertainties expressed as the standard deviation. If the hypothesized scaling of the
Lyapunov exponent on Kolmogorov time scale were correct, these data
would, within their uncertainty, fall along a horizontal
line. However, this does not appear to be the case. Indeed,
$\bar\gamma\tau_\eta$ appears to be growing with $Re_\lambda$. Also,
shown in figure~\ref{fig:dependence} is the dependence of scaled
Lyapunov exponent on domain size at constant Reynolds number. These
data do appear to be consistent with the hypothesis that the Lyapunov
exponent does not depend on the domain size.

To make these scaling observations quantitative, Bayesian inference
is used to infer the coefficients $\alpha$ and $\beta$ in a scaling
relationships of the form 
\begin{equation}
\bar\gamma\tau_\eta=\alpha_1 Re_\lambda^{\beta_1}\quad\mbox{and}\quad
\bar\gamma\tau_\eta=\alpha_2 (\mathcal{L}/L)^{\beta_2},
\label{eq:models}
\end{equation}
given the data
and its uncertainties. These scaling relationships serve as the
``model'' for the inference. In Bayesian inference for this problem, the
joint probability distribution $\pi(\alpha,\beta|{\bf d})$ of the
parameters $\alpha$ and $\beta$ conditioned on data ${\bf d}$
(shown in table~\ref{tab:cases}) is sought. Bayes' theorem gives this conditional
probability as:
\begin{equation}
  \pi(\alpha,\beta|{\bf d})\propto \pi({\bf
    d}|\alpha,\beta)\pi(\alpha,\beta)
\end{equation}
where $\pi({\bf d}|\alpha,\beta)$ is the likelihood and
$\pi(\alpha,\beta)$ is the prior. The likelihood is the joint
probability density for the observed quantities evaluated for the
observed values of these quantities, as determined by the model with
parameters $\alpha$ and $\beta$, and given the uncertainties in the
data. The prior represents our prior knowledge about the parameters,
independent of the data.

The data are statistical averages obtained from direct numerical
simulations. The primary source of uncertainty in such data is
statistical sampling error. The central limit theorem implies that in
the limit of large samples, the uncertainty associated with sampling
error is normally distributed with zero mean. Therefore, to formulate the
likelihood, the data are assumed to have Gaussian uncertainty with
standard deviations as reported in table~\ref{tab:data}. The probability
distribution for the $i$th observation of the value of $\bar\gamma$ as predicted by the models
is thus given by
\begin{equation}
  \pi(\bar\gamma|\alpha,\beta,x_i)=\frac{1}{\sigma_{\gamma_i}\sqrt{2\pi}}\exp\left[-\frac{(\bar\gamma-\alpha_1x_i^{\beta_1})^2}{2\sigma_{\gamma_i}^2}\right]
\end{equation}
where $x_i$ is the independent variable ($Re_{\lambda i}$ or
$\mathcal{L}_i/L$, depending on which scaling relation is being
inferred) of the $i$th observation and $\sigma_{\gamma_i}$ is the
standard deviation in $\bar\gamma$ associated with the $i$th
observation. However, there are also uncertainties in the values of
the independent variables $x$, as determined from the DNS, again with
a Gaussian distribution and standard deviation for the $i$th
observation of $\sigma_{x_i}$. In this case, the probability
distribution of the independent variable $x$ given the observation $x_i$
is
\begin{equation}
  \pi(x|x_i)=\frac{1}{\sigma_{x_i}\sqrt{2\pi}}\exp\left[-\frac{(x-x_i)^2}{2\sigma_{x_i}^2}\right].
\end{equation}
The conditional distribution of $\bar\gamma$ given the parameters and
the observed independent variable is then
given by
\begin{equation}
  \pi(\bar\gamma|\alpha,\beta,x_i)=\int_x\pi(\gamma|\alpha,\beta,x)\pi(x|x_i)\,
  dx.
  \label{eq:distgamma}
\end{equation}
Finally, to obtain the likelihood, (\ref{eq:distgamma}) is evaluated
at $\bar\gamma=\gamma_i$ and the uncertainties in each observation are
assumed to be independent (an excellent assumption), yielding:
\begin{equation}
  \pi({\bf d}|\alpha,\beta)=\prod_i
  \pi(\bar\gamma=\gamma_i|\alpha,\beta,x_i).
\end{equation}

To inform the prior, we consider the range of time scales in the
turbulence. The largest is the eddy turn-over time, which is
proportional to $q^2/\epsilon$, and the smallest is the Kolmogorov
time scale $\sqrt{\nu/\epsilon}$. The ratio of the turnover to the
Kolmogorov times scales as $Re_\lambda$. Therefore, the Lyapunov
exponent $\bar\gamma\tau_\eta$ scaling with the turn-over time would
imply $\beta=-1$ and scaling with the Kolmogorov scale would imply
$\beta=0$. However, theoretical arguments suggest that the Lyapunov
exponent scales with the Kolmogorov time scale \cite{crisanti1993intermittency} ($\beta=0$),
and we need to allow for the possibility that this assessment may be
in error in either direction. The bounds on the range of plausible
values of $\beta$ were therefore extended to $-1\le\beta\le1$, and a
uniform distribution over this range was used as a prior for
$\beta$. Somewhat arbitrarily, the same range was used for the $\beta$
prior in the domain size scaling relationship. The parameter $\alpha$ is a positive definite scaling
parameter, and so following Jaynes \cite{Jaynes_2003}, a Jeffries distribution
$\pi(\alpha)\sim1/\alpha$ is used as an (improper) prior. Finally, the
priors for $\alpha$ and $\beta$ are independent so
$\pi(\alpha,\beta)=\pi(\alpha)\pi(\beta)$.

\begin{figure}[t]
\includegraphics[width=\linewidth]{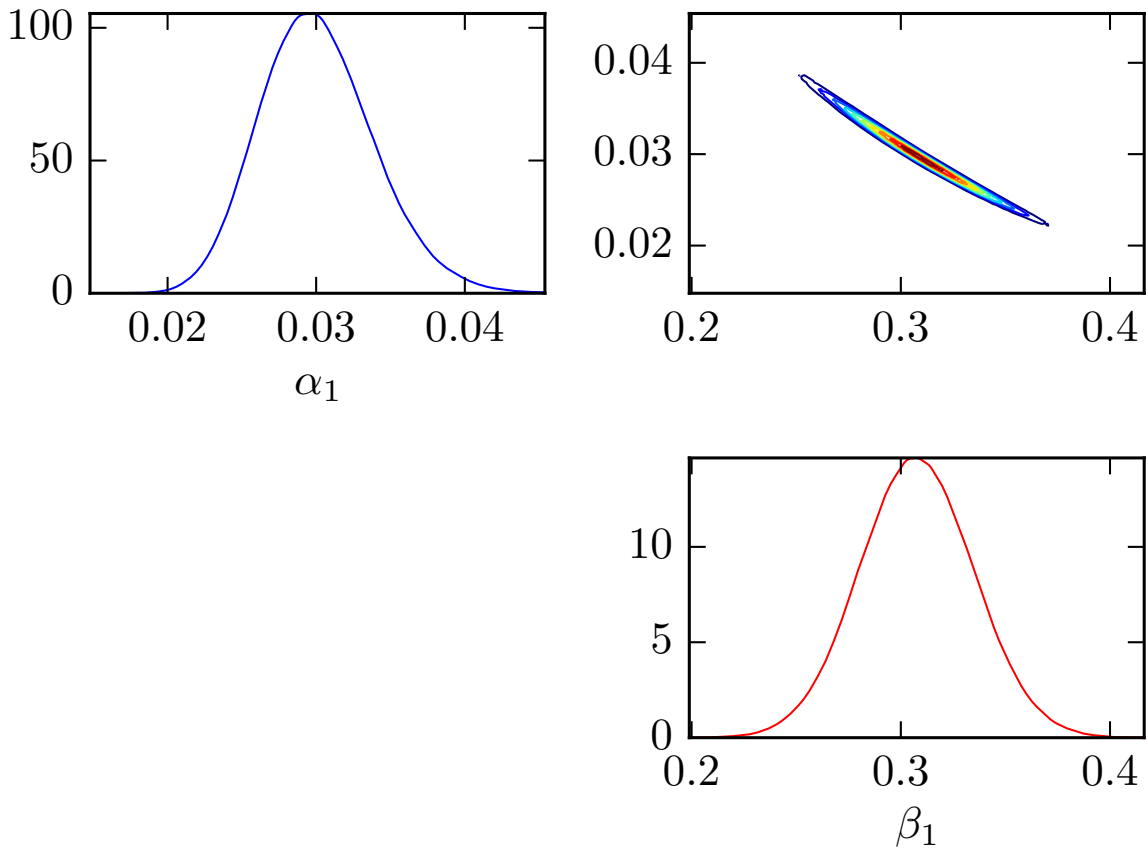}
\caption{Posterior PDFs for $\alpha_1$ and $\beta_1$ in the Reynolds
number scaling model (\ref{eq:models}). Shown are the marginal
distributions of both parameters along with contours of their joint distribution. }
  \label{fig:rescaling}
\end{figure}
\begin{figure}[t]
\includegraphics[width=\linewidth]{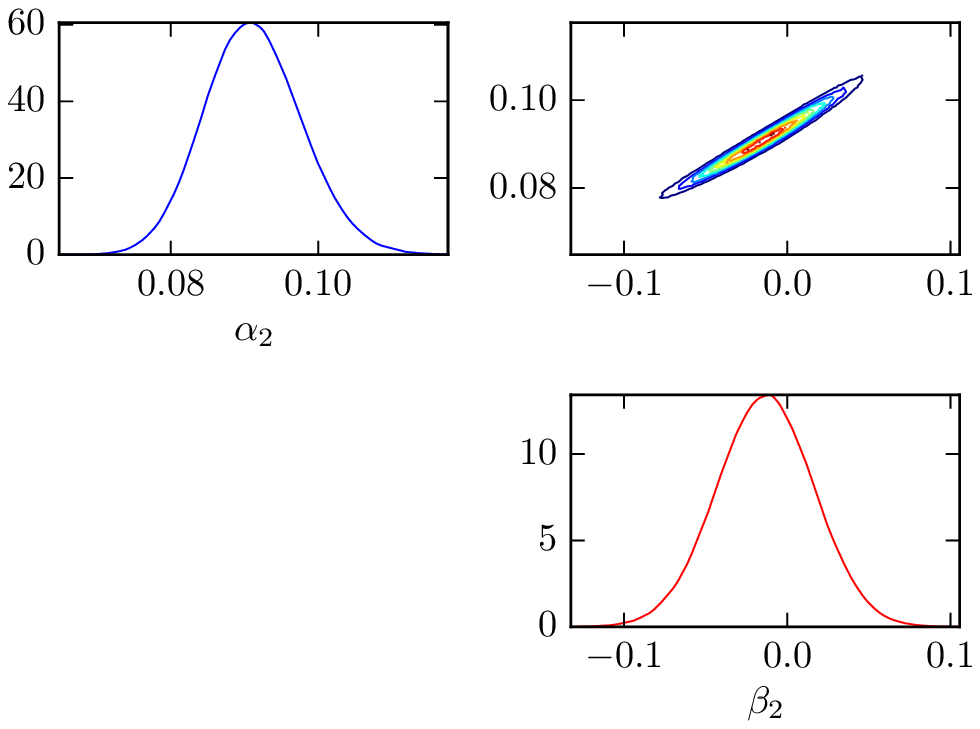}
\caption{Posterior PDFs for $\alpha_2$ and $\beta_2$ in the domain size
scaling model (\ref{eq:models}). Shown are the marginal
distributions of both parameters along with contours of their joint distribution.}
\label{fig:Bscaling}
\end{figure}

Given the likelihood and prior described above, and the data in
Table~\ref{tab:data}, samples of the posterior distribution were
obtained using a Markov-chain Monte Carlo (MCMC) algorithm \cite{haario2006dram}
as implemented in the QUESO library \cite{Prudencio2012, mcdougall2015parallel}. The resulting samples
were used to characterize the joint posterior distribution of $\alpha$
and $\beta$ for both the Reynolds number and domain size scaling model,
as shown in figures~\ref{fig:rescaling} and
\ref{fig:Bscaling}. Notice in these figures that the joint
distribution has probability mass concentrated in thin diagonally
oriented regions, showing that uncertainty in $\alpha$ and $\beta$
are highly correlated. Indeed, the uncertainty in the $\beta$'s is as
large as it is because changes in $\beta$ can be compensated for by
changes in $\alpha$ so that the model still fits the data.
The MCMC samples were also used to determine the
uncertainty in the model predictions for the Lyapunov exponent as a
function of Reynolds number and domain size, with the results plotted in
figure~\ref{fig:dependence}, along with the data. From this, it is clear that
the scaling models as calibrated are consistent with the data and
their uncertainty.

The marginal posterior distribution for $\beta$ in the Reynolds number
scaling relation shows that the most likely values of $\beta$ are
between about $1/4$ and $1/3$, with the possibility that the value is
zero essentially precluded. This is remarkable since it suggests
instability time scales that will become increasingly faster than
Kolmogorov with increasing Reynolds number. The origin of this fast
time scale is currently unclear. One possibility to consider is that
this fast instability time scale arises as an artifact of the time
discretization of the DNS. However the DNS time step in Kolmogorov
units $\Delta t/\tau_\eta\sim Re_\lambda^{-1/2}$, so if the Lyapunov
exponent were scaling with the DNS time step, $\beta$ would be 1/2,
which is also essentially precluded by the posterior distribution. The
time discretization thus appears to be an unlikely origin of the
observed Reynolds number scaling. This was also verified by running
a time refinement study where $\bar\gamma$ was found to be invariant
to changing $\Delta t$.

As with the time step, interest in the computational domain size
arises because of concern that computational artifacts not impact our
Lyapunov exponent analysis. The posterior distribution of $\beta$ in
the domain size scaling relationship (figure~\ref{fig:Bscaling}) shows that
$\beta=0$ is highly likely, with the most probable values of $\beta$
ranging from -0.05 to 0.05. If there is an effect of the domain
size, the data indicates that it is extremely weak. It therefore
appears that the Lyapunov exponent Reynolds number scaling discussed
above and the short-time Lyapunov exponent analysis presented in
section~\ref{sec:short} are unaffected by finite domain size effects.

\section{Short-Time Lyapunov Exponent Analysis}
\label{sec:short}
As discussed in section~\ref{sec:lyapunov}, both the disturbance field
($\delta u$) used to compute the Lyapunov exponent and its
instantaneous exponential growth rate ($\gamma'$) depend only on the
instantaneous Navier-Stokes velocity $u$, not on the initial
disturbance. In short-time Lyapunov exponent analysis, we study $\gamma'$ and
$\delta u$ to learn about the instabilities responsible for the chaotic
nature of turbulence.

\begin{figure}[t]
\begin{center}
    \subfigure[Case 1, $Re_\lambda=37$] {\includegraphics[width=0.5\linewidth]{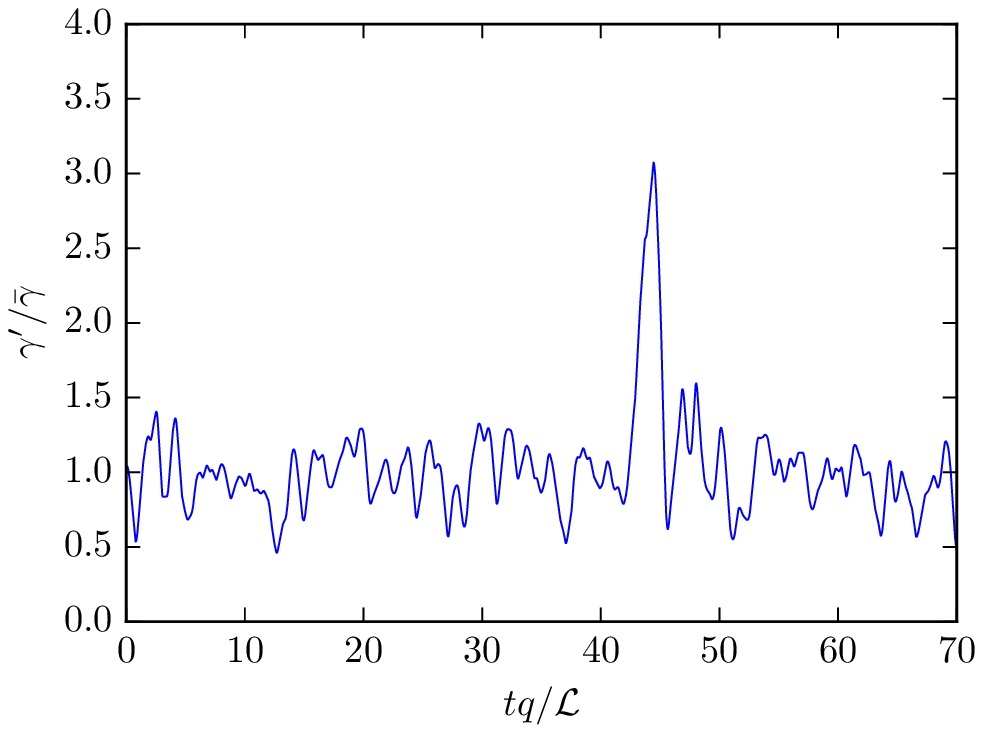}}%
    \subfigure[Case 8, $Re_\lambda=37$, large $L/\mathcal{L}$] {\includegraphics[width=0.5\linewidth]{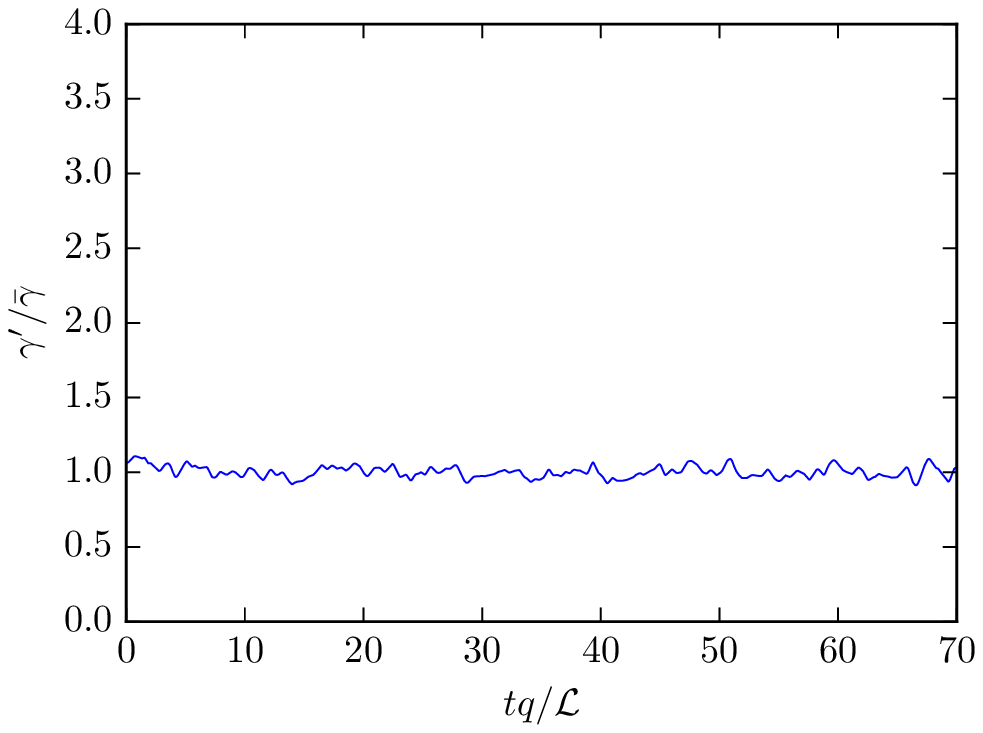}}
    \subfigure[Case 6, $Re_\lambda=210$] {\includegraphics[width=0.5\linewidth]{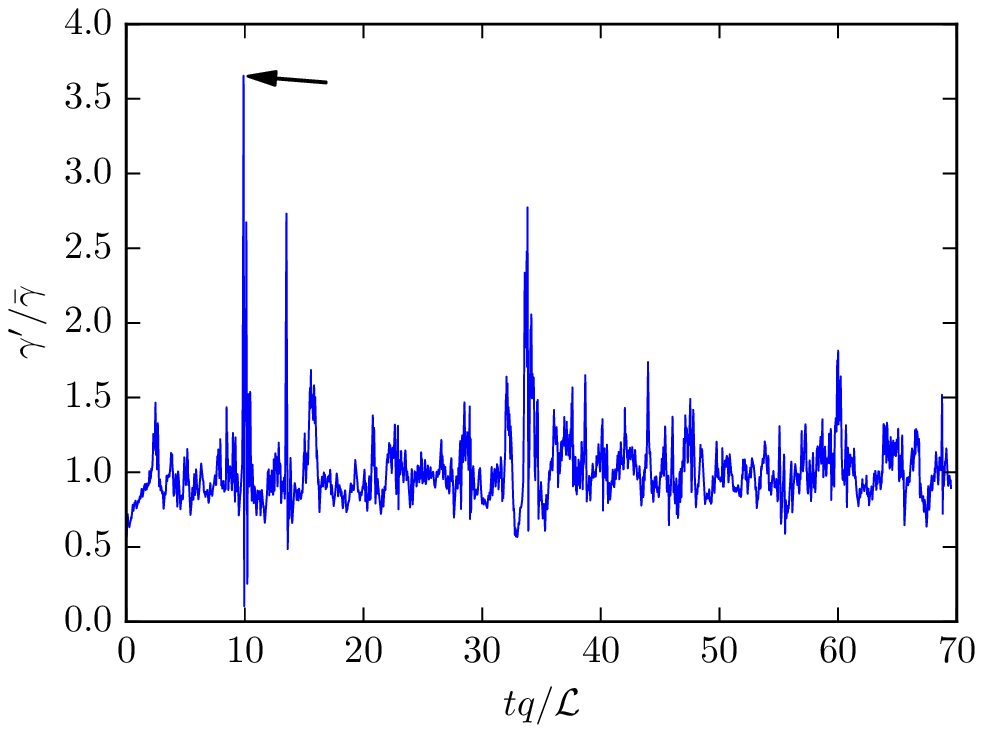}}%
    \subfigure[Case 6, $Re_\lambda=210$ (zoomed in)] {\includegraphics[width=0.5\linewidth]{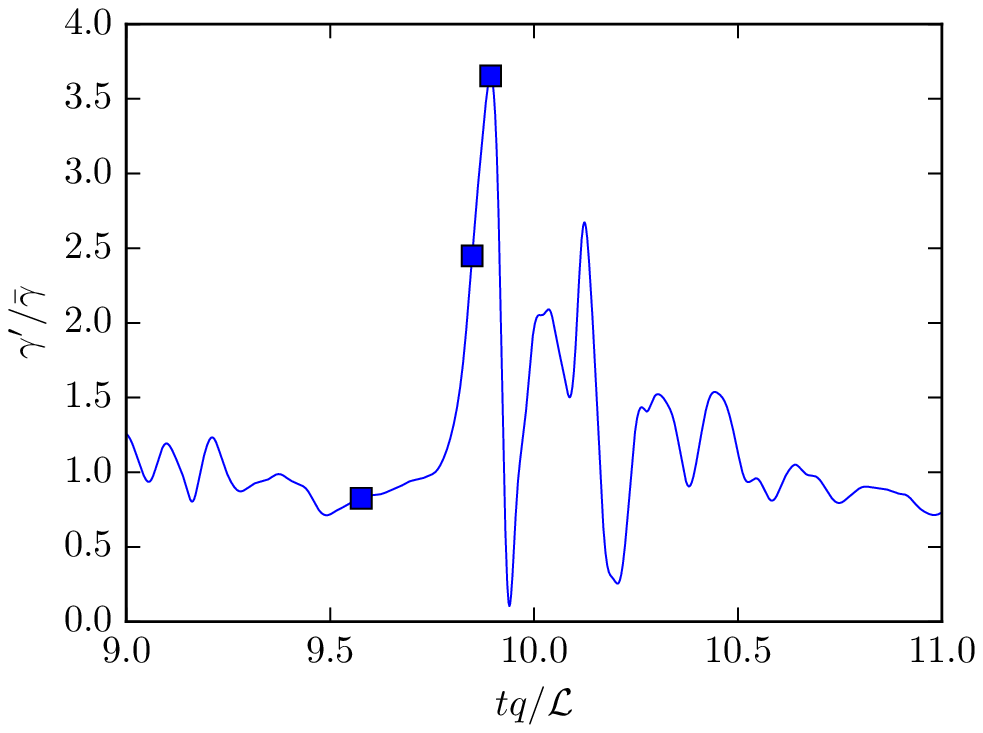}}
\end{center}
\caption{Short-time Lyapunov exponent scaled by $\bar\gamma$. In (d),
  the time axis is expanded to zoom in on the peak indicated in
  (c), and symbols show the times at which the images in
  figure~\ref{fig:peak-field} were obtained.  }
\label{fig:lyapunov-time}
\end{figure}

\begin{figure}[t]
\begin{center}
    \subfigure[Case 1, $Re_\lambda=37$]  {\includegraphics[width=0.5\linewidth]{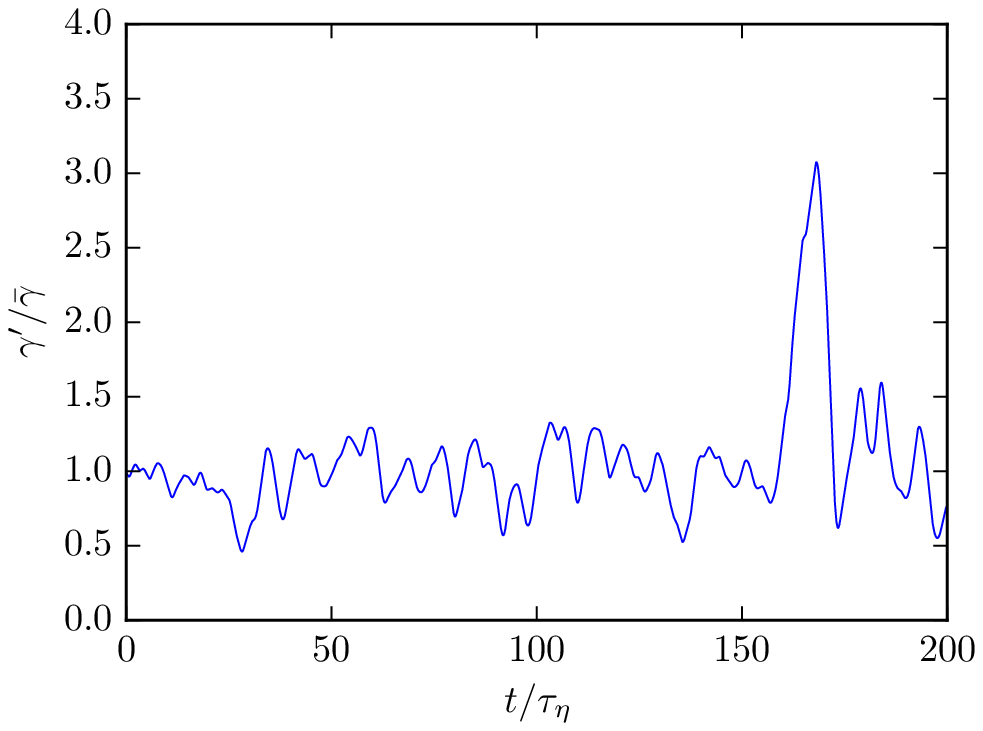}}%
    \subfigure[Case 6, $Re_\lambda=210$] {\includegraphics[width=0.5\linewidth]{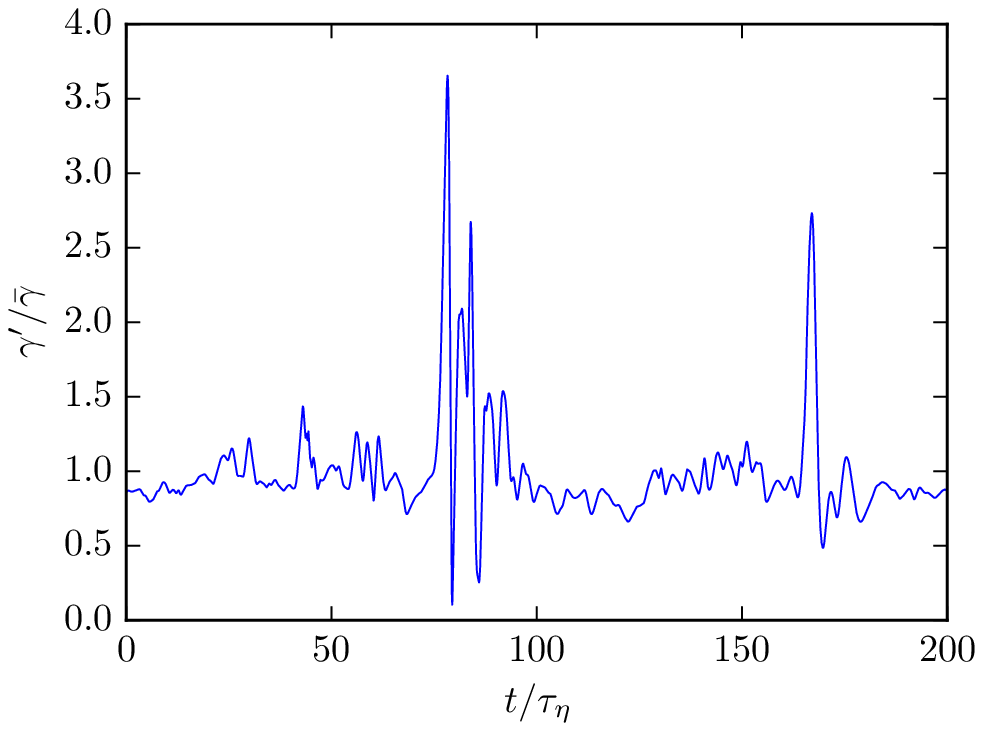}}%
\end{center}
\caption{Short-time Lyapunov exponent $\gamma'/\bar\gamma$ with time
  scaled by $\tau_\eta$. 
}
\label{fig:lyapunov-time-ktime}
\end{figure}

First, consider the time evolution of the exponential growth rate $\gamma'$,
which is shown in figure~\ref{fig:lyapunov-time} for $Re_\lambda=37$
and 210 (cases 1 and 6 respectively), normalized by $\bar\gamma$. Note
that in both cases $\gamma'$ takes large excursions from the mean, of
order 3 times the mean value. However, the variations in $\gamma'$
occur on a much shorter time scale and the large excursions seem to
occur more often in the high Reynolds number case. The
time scale on which $\gamma'$ varies appears to decrease somewhat faster
than the Kolmogorov time scale with increasing Reynolds number, as when plotted against $t/\tau_\eta$,
$\gamma'$ still varies faster for case 6
(figure~\ref{fig:lyapunov-time-ktime}).  At the same Reynolds number
(cases 1 and case 8), the variability of $\gamma'$ decreases sharply
with increasing relative computational domain size $L/\mathcal{L}$. The fact
that the time scale of the instability, as measured by the Lyapunov
exponent, decreases faster than the Kolmogorov time scale suggests
that the instability processes are acting at spatial scales near the
Kolmogorov scale. In this case, a simulation with a larger domain size
relative to intrinsic turbulence length scales would include a larger
sample of local unstable turbulent flow features, resulting in smaller
variability in $\gamma'$. In comparing case 8 with case 1, the
relative volume increases by a factor 64, suggesting that the
variability of $\gamma'$ should be about a factor of 8 smaller in case
8 than in case 1, which is indeed consistent with the data.

\begin{figure}[t]
    \subfigure[$tq/\mathcal{L}=9.58$] {\includegraphics[bb=478 194 1220 937,clip,width=0.3\linewidth]{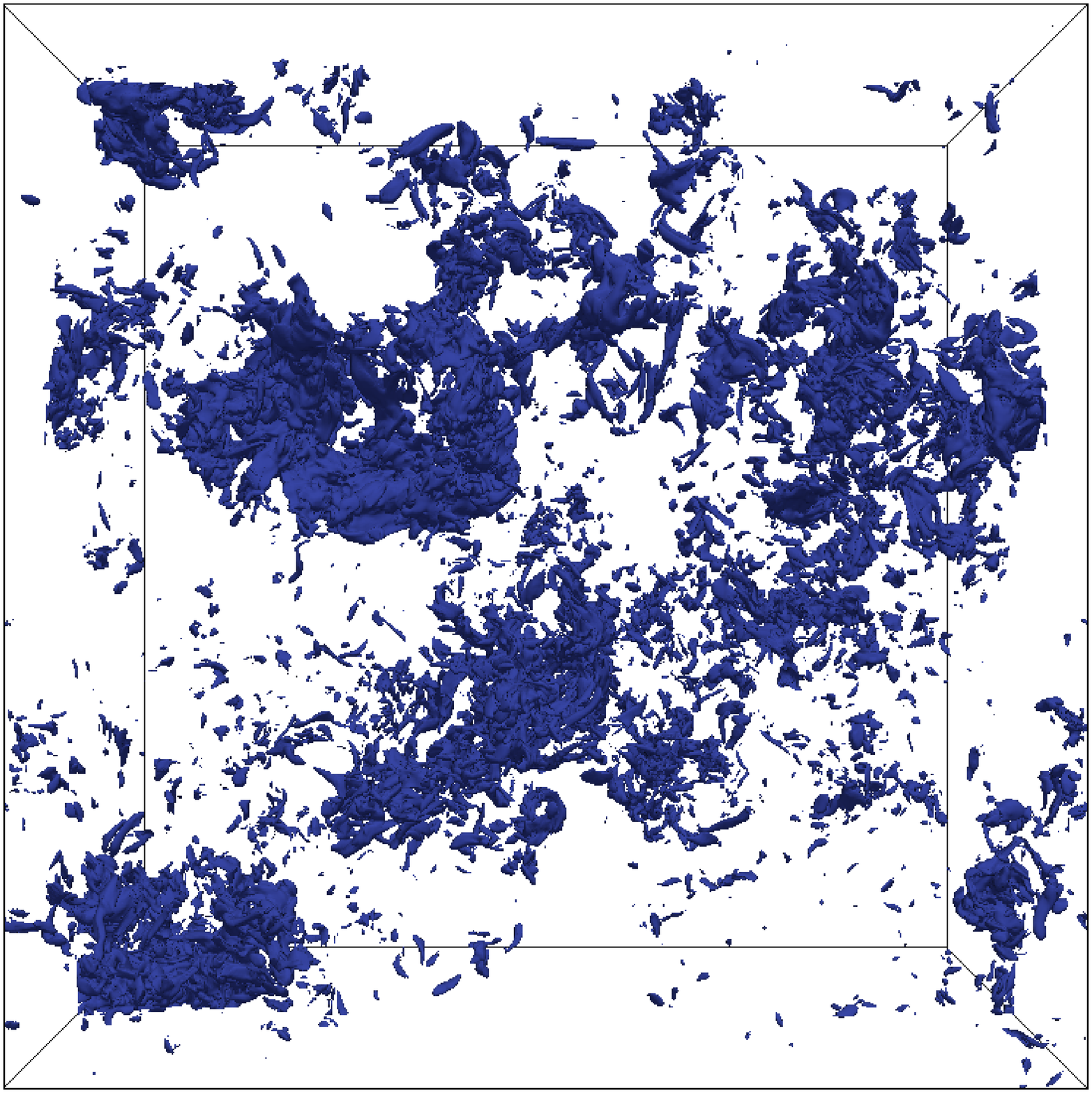}}\hfill%
    \subfigure[$tq/\mathcal{L}=9.85$] {\includegraphics[bb=478 194 1220 937,clip,width=0.3\linewidth]{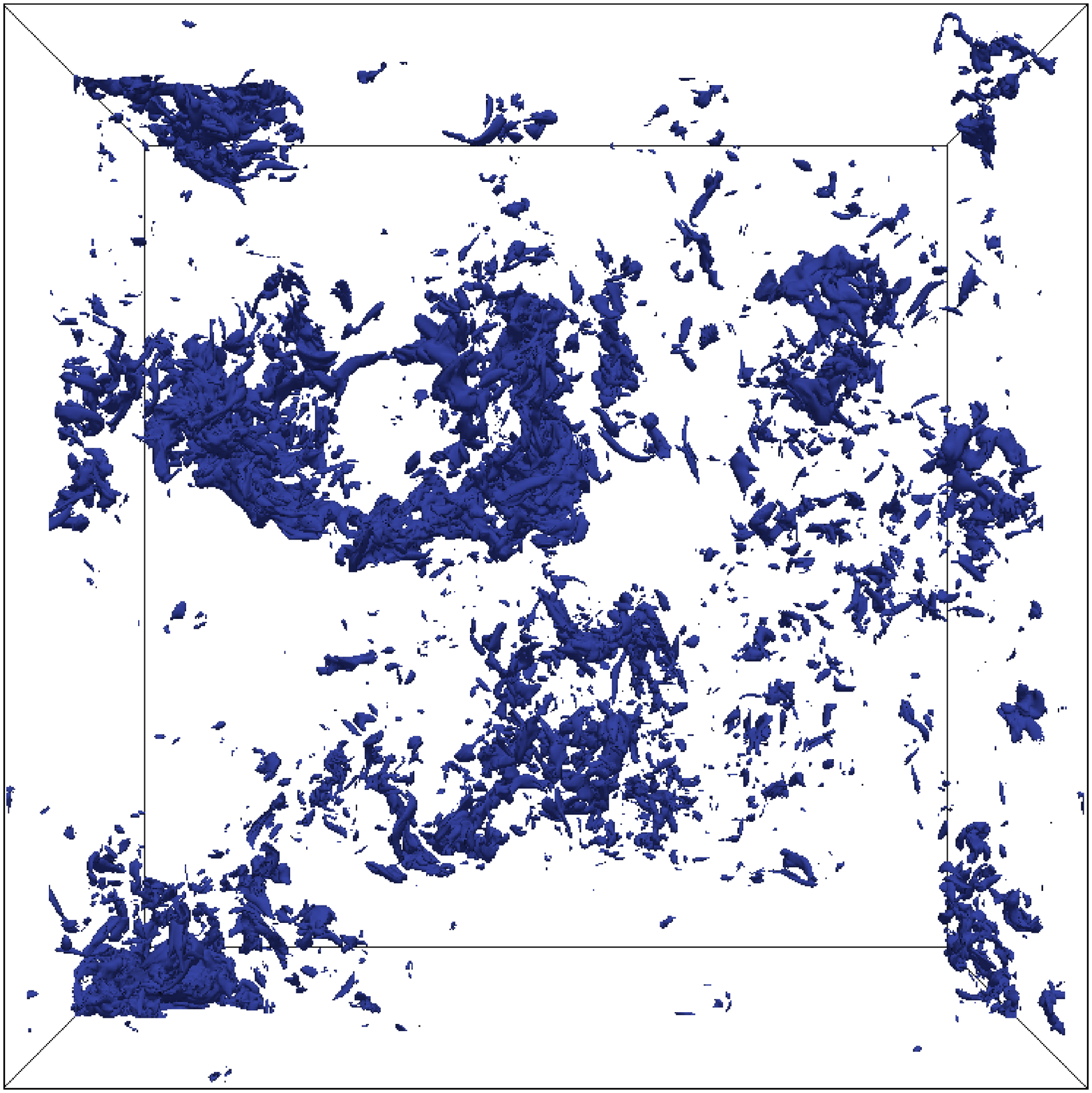}}\hfill%
    \subfigure[$tq/\mathcal{L}=9.89$] {\includegraphics[bb=478 194 1220 937,clip,width=0.3\linewidth]{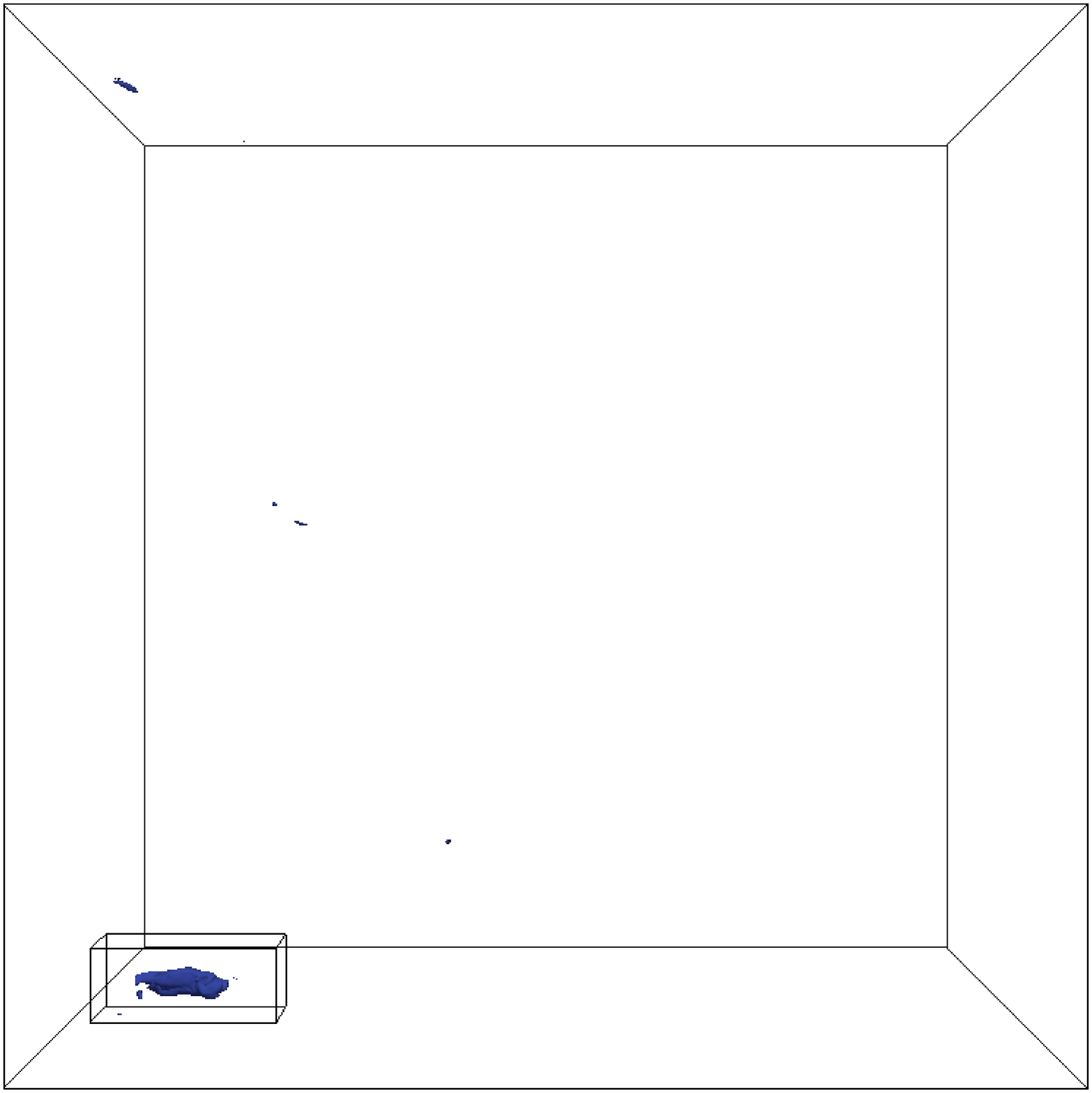}}
\caption{Contour of the magnitude of $\delta u$ at three times leading
  up to the peak as indicated in figure~\ref{fig:lyapunov-time}
  for $Re_\lambda=210$ (case 6). The
  contour shown is that at which 60\% of the disturbance energy is
  enclosed by the contour. To achieve this, the contour levels are (a)
  15, (b) 25, and (c) 2500 times the mean disturbance energy density.}
\label{fig:peak-field}
\end{figure}

At the peaks in $\gamma'$, the growth of the disturbance energy is
particularly rapid, and the question naturally arises as to what is
special about these times. To investigate this, the spatial
distribution of the magnitude of the disturbance energy density is
visualized in figure~\ref{fig:peak-field} at three times, just before
the beginning of a peak in $\gamma'$, a time half way up that peak and
at the peak ($tq/\mathcal{L}=9.58$, 9.85 and 9.89 in
figure~\ref{fig:lyapunov-time}). Notice that before the rapid growth
of $\gamma'$ into the peak, the energy in the disturbance field is
broadly distributed across the spatial domain. Half way up the peak,
the distribution is much more spotty, and finally at the peak, the
disturbance energy is primarily focused in a small region, appearing
in the lower left corner of figure~\ref{fig:peak-field}(c). The
contour levels in these images were chosen so that the contours
enclose 60\% of the disturbance energy, implying that 60\% of the
disturbance energy is concentrated in the small feature in the lower
left of figure~\ref{fig:peak-field}(c). Another indication of the
dominance of the disturbance feature in figure~\ref{fig:peak-field}(c)
is that the contour level needed to enclose 60\% of the energy is about
2500 times the mean disturbance energy density, while in
figure~\ref{fig:peak-field}(a) the contour is only about 15 times the
mean. Clearly the growth of the disturbance field in this concentrated
area is responsible for the peak in $\gamma'$. However, the spatially
local exponential growth rate of the disturbance energy $|\delta
u|^{-2}\,\partial|\delta u|^2/\partial t$ is not particularly large
there, large values of this quantity are distributed broadly
across the spatial domain. It seems, then, that the large peak
in $\gamma'$ is due to a local disturbance that is able to grow over
an extended time until it dominates the disturbance energy, so that
the disturbance is localized in a region of relatively large growth
rate. This is presumably unusual because it requires that the local
unstable flow structure responsible for the disturbance growth
persists for a long time.

\begin{figure}[t]
\begin{center}
\includegraphics[width=0.6\linewidth]{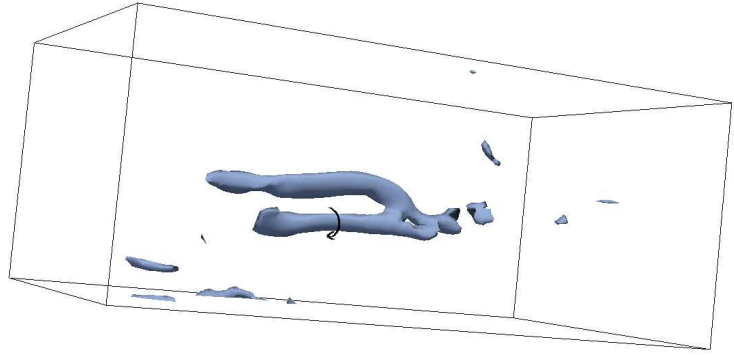}
\end{center}
\caption{Contour of the magnitude of the vorticity of the base field
  for $Re_\lambda=210$ (case 6) at the peak of $\gamma'$ in the region
  where the disturbance field is localized (box highlighted in figure~\ref{fig:peak-field}c) . The contour level is
  9.2 times the square root of the mean enstrophy. The vortex tubes are co-rotating, with the direction of rotation indicated by the black arrow.}
\label{fig:base-vorticity}
\end{figure}

\begin{figure}[t]
\begin{center}
\includegraphics[width=0.6\linewidth]{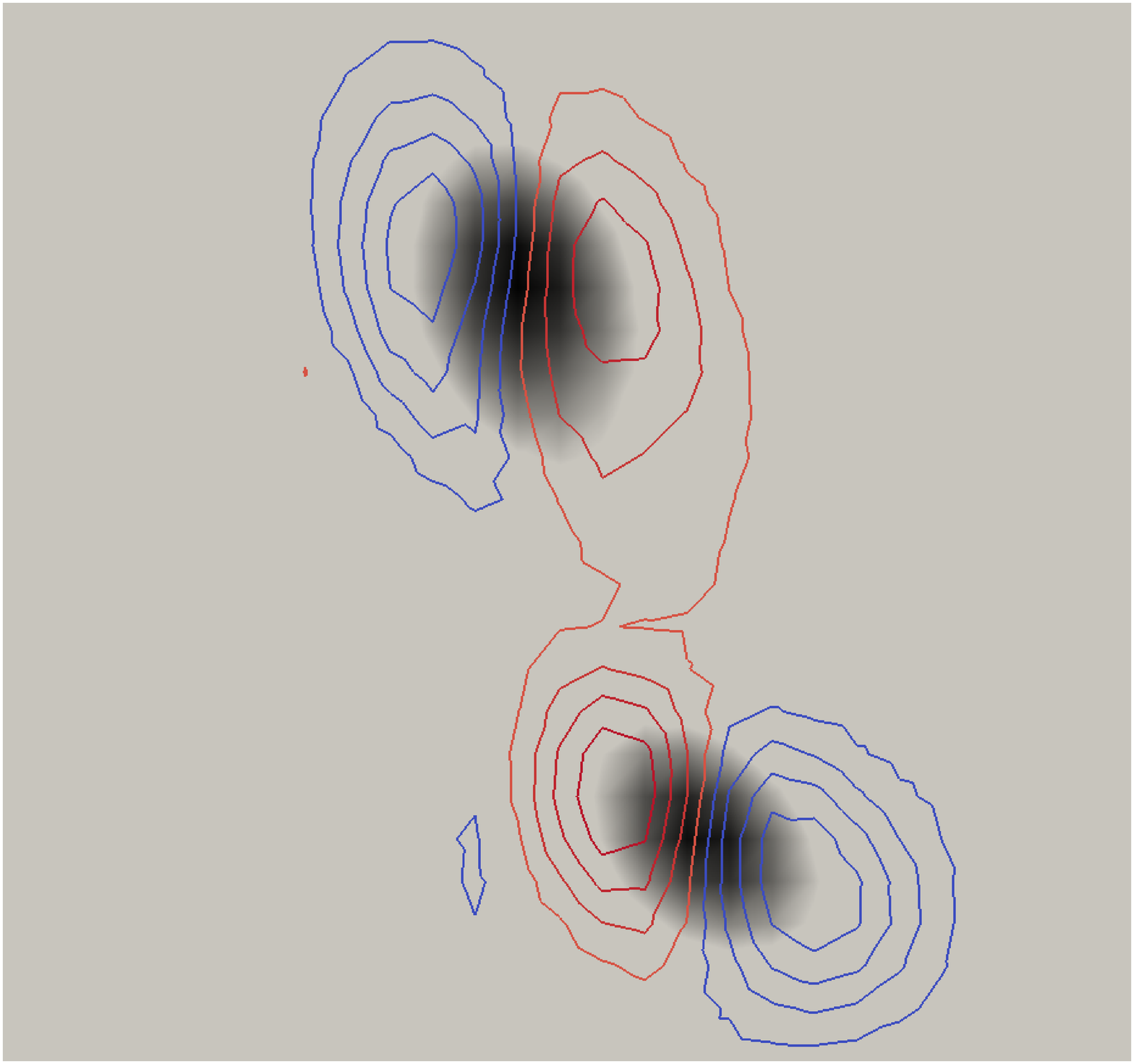}
\end{center}
\caption{The magnitude of the vorticity (grayscale) and the disturbance
  vorticity component normal to the plane (contour lines)
  in a plane perpendicular to and in the
  middle of the vortex tubes shown in figure~\ref{fig:base-vorticity}.
  For the disturbance vorticity, the red and blue contours are of opposite signs.}
\label{fig:slice-vorticity}
\end{figure}

It is of interest to investigate the turbulent flow structures
responsible for the large localized disturbance energy at the peak in
$\gamma'$. In the region where $\delta u$ is localized, the base
field exhibits a pair of co-rotating vortex tubes
(figure~\ref{fig:base-vorticity}). As shown in
figure~\ref{fig:slice-vorticity}, the disturbance vorticity is
localized on the vortex tubes, with regions of opposite signed
disturbance vorticity to one side or the other of each vortex
tube. This disturbance, when added to the base field would have the
effect of displacing each vortex tube along the line between the
positive and negative peaks in the disturbance vorticity associated
with each tube. The instability then appears to be one associated with
slowing (speeding up) the co-rotation of the vortex tubes while they
move away from (toward) each other. Note that the disturbance equations,
being linear and homogeneous, are invariant to a sign change, and so
the sign of the vortex displacement is indeterminate. Such an instability
of co-rotating vortices is reminiscent of the pairing instability in
two-dimensional mixing layers.

\section{Conclusions}
\label{sec:conclude}
The results of the scaling study (section~\ref{sec:scaling}) show
definitively that, at least over the Reynolds number range studied,
the Lyapunov exponent does not scale like the inverse Kolmogorov time
scale, as had been previously suggested
\cite{aurell1996growth}. Instead, $\bar\gamma\tau_\eta$ increases with
Reynolds number like $Re_\lambda^\beta$ for $\beta$ in the range from
1/4 to 1/3. Further note that the analysis of Aurell {\it et al.~}\cite{aurell1996growth}
indicated that a correction for the
intermittance of dissipation would yield $\beta < 0$, also
inconsistent with the current results.  If positive $\beta$ scaling
holds to much higher Reynolds numbers, it would be remarkable, as it
would mean that there are instability processes that act on time
scales shorter than Kolmogorov. However, in the highest Reynolds
number ($Re_\lambda=210$) simulation performed here,
$\bar\gamma\tau_\eta$ is still only 0.16. It is certainly possible
that this Reynolds number dependence of $\bar\gamma\tau_\eta$ is a low
Reynolds number effect, caused by insufficient scale separation
between the large scales and the scales at which the instabilities
act, and that the value will reach a plateau at some much higher
Reynolds number.  Clearly, this scaling behavior of the maximum
Lyapunov exponent is worthy of further study. The current results
suggest that the generally accepted and most obvious scaling is not
correct, and that, unfortunately, turbulent fluctuations are even less
predictable than previously thought.

The short-time analysis described in section~\ref{sec:short} confirmed
that the dominant instabilities in turbulence act on the smallest
eddies. Further, at $Re_\lambda=210$, when the instantaneous
disturbance growth rate was the largest (about 3 times the mean), the
disturbance energy was highly localized, suggesting that it was a
particular local instability that was responsible for the rapid growth
at that time. However, this was not due to a particularly large local
growth rate, as the logarithmic time derivative of the spatially local
disturbance energy was equally large in regions spread throughout the
domain. It may be that the localized instability we observed is not of
particular importance, except that the underlying structure in the
turbulent field was especially long-lived. None-the-less, studying it
showed that one of the possible instability mechanisms acting in
turbulence is reminiscent of pairing instabilities of co-rotating
vortices, as in a mixing layer. In this, the short-time Lyapunov
analysis pursued here appears to be a valuable tool for the study of the
instabilities underlying turbulence.

\section*{Acknowledgements} Support for the research reported here was
provided by the Department of Energy through the Center for Exascale
Simulation of Combustion in Turbulence (ExaCT) under subcontract to
Sandia National Laboratory, project 1174449, and is gratefully
acknowledged. We also wish to thank Dr.~Myoungkyu Lee for his assistance with
code development for the simulations.

\bibliographystyle{abbrvnat}
\bibliography{paper}
\end{document}